\documentclass[a4paper,fleqn]{cas-dc}

\usepackage[authoryear]{natbib}
\usepackage{amsmath}
\usepackage[ruled,vlined]{algorithm2e}
\usepackage{caption}
\usepackage{subfigure}
\usepackage{multirow}
\usepackage{booktabs}
\usepackage{eqparbox}
\usepackage[figuresright]{rotating}

\def\tsc#1{\csdef{#1}{\textsc{\lowercase{#1}}\xspace}}
\tsc{WGM}
\tsc{QE}
\tsc{EP}
\tsc{PMS}
\tsc{BEC}
\tsc{DE}

\begin{document}
\sloppy{}

\let\WriteBookmarks\relax
\def\floatpagepagefraction{1}
\def\textpagefraction{.001}
\shorttitle{Efficient and Effective Local Search for the SUKP and BMCP}
\shortauthors{Zhu et~al.}

\author[1]{Wenli Zhu}
\author[2]{Liangqing Luo}

\address[1]{School of Statistics, Jiangxi University of Finance and Economics, Nanchang Jiangxi 330013, China; Email: zzhuwenli@163.com}

\address[2]{School of Statistics, Jiangxi University of Finance and Economics, Nanchang Jiangxi 330013, China; Corresponding author, Email: llq6429@163.com}

\title [mode = title]{Efficient and Effective Local Search for the Set-Union Knapsack Problem and Budgeted Maximum Coverage Problem} 



\begin{abstract}
The Set-Union Knapsack Problem (SUKP) and Budgeted Maximum Coverage Problem (BMCP) are two closely related variant problems of the popular knapsack problem. Given a set of weighted elements and a set of items with nonnegative values, where each item covers several distinct elements, these two problems both aim to find a subset of items that maximizes an objective function while satisfying a knapsack capacity (budget) constraint. We propose an efficient and effective local search algorithm called E2LS for these two problems. To our knowledge, this is the first time that an algorithm has been proposed for both of them. E2LS trade-offs the search region and search efficiency by applying a proposed novel operator ADD$^*$ to traverse the refined search region. Such a trade-off mechanism allows E2LS to explore the solution space widely and quickly. The tabu search method is also applied in E2LS to help the algorithm escape from local optima. Extensive experiments on a total of 168 public instances with various scales demonstrate the excellent performance of the proposed algorithm for both the SUKP and BMCP.
\end{abstract}

\begin{keywords}
Set-union knapsack problem \sep Budgeted maximum coverage problem \sep Local search \sep Tabu search
\end{keywords}

\maketitle

\section{Introduction}
\label{Sec_Intro}

The Set-Union Knapsack Problem (SUKP)~\citep{Goldschmidt1994} and Budgeted Maximum Coverage Problem (BMCP)~\citep{Khuller1999} are two closely related NP-hard combinatorial optimization problems. Let $I = \{i_1,...,i_m\}$ be a set of $m$ items where each item $i_j, j \in \{1,...,m\}$ has a nonnegative value $v_j$, $E = \{e_1,...,e_n\}$ be a set of $n$ elements where each element $e_k, k \in \{1,...,n\}$ has a nonnegative weight $w_k$, and $C$ is the capacity of a given knapsack in the SUKP (or the budget in the BMCP). The items and elements are associated by a relation matrix $R \in \{0,1\}^{m \times n}$, where $R_{jk} = 1$ indicates that $e_k$ is covered by $i_j$, otherwise $R_{jk} = 0$. The SUKP aims to find a subset $S$ of $I$ that maximizes the total value of the items in $S$, at the same time the total weight of the elements covered by the items in $S$ does not exceed the capacity $C$. The SUKP can be stated formally as follows. 

\begin{equation}
\label{eq_f}
    \text{Maximize}~~f(S) = \sum\nolimits_{j \in \{j | i_j\in S\}}v_j,
\end{equation}
\vspace{-1.5em}
\begin{equation}
\label{eq_W}
    \text{Subject to}~~W(S) = \sum\nolimits_{k \in \{k | R_{jk} = 1, i_j \in S\}}w_k \leq C.
\end{equation}

For the BMCP, the goal is to find a subset $S$ of $I$ that maximizes the total weight of the elements covered by the items in $S$, while the total value of the items in $S$ does not exceed the capacity (budget) $C$. The BMCP can be stated formally as follows. 

\begin{equation}
    \text{Maximize}~~W(S),
\end{equation}
\vspace{-1.5em}
\begin{equation}
    \text{Subject to}~~f(S) \leq C.
\end{equation}

Obviously, the SUKP and BMCP can be transferred to each other by swapping the optimization objective and constraint objective. Both the SUKP and BMCP are computationally challenging and have many real-world applications, such as flexible manufacturing~\citep{Goldschmidt1994}, financial decision making~\citep{Khuller1999,Kellerer2004}, data compression~\citep{Yang2016}, software defined network optimization~\citep{Kar2016}, project investment~\citep{Wei2019}, etc.

Exact and approximation algorithms are two kinds of methods for the SUKP and BMCP. For example, exact algorithms for the SUKP based on dynamic programming~\citep{Arulselvan2014} and linear integer programming~\citep{Wei2019}, and greedy approximation algorithms for the SUKP~\citep{Taylor2016} and BMCP~\citep{Khuller1999}. These algorithms can all theoretically guarantee the quality of their solutions, but the exact algorithms are hard to scale to large instances, and the approximation algorithms are hard to yield high-quality results.

Heuristic algorithms such as population-based algorithms and local search algorithms are more practical than exact and approximation algorithms. Population-based methods usually use bio-inspired metaheuristic operations among the population, so as to find excellent individuals. He et al.~\citep{He2018} first proposed a binary artificial bee colony algorithm for the SUKP. Later on, a swarm intelligence-based algorithm~\citep{Ozsoydan2019} was proposed to solve the SUKP. There are also some hybrid algorithms for the SUKP that combine population-based methods with local search to improve the performance. For example, Lin et al.~\citep{Lin2019} combined binary particle swarm optimization with tabu search, Dahmani et al.~\citep{Dahmani2020} combined swarm optimization with local search operators, Wu and He~\citep{Wu2020} proposed a hybrid Jaya algorithm for the SUKP. Population-based methods have attracted lots of attention. However, these methods are more complex to design than local search algorithms, and the performance of existing population-based methods is not as good as the state-of-the-art local search methods for the SUKP.

For the local search methods, the I2PLS algorithm~\citep{Wei2019} is the first local search method for the SUKP, which is based on tabu search. Later on, some better tabu search methods were proposed~\citep{Wei2021KBTS,Wei2021MSBTS}. Wei and Hao~\citep{Wei2021MSBTS} tested these tabu search algorithms on SUKP instances with no more than 1,000 items and elements. Recently, Zhou et al.~\citep{Zhou2021} proposed an efficient local search algorithm called ATS-DLA for the SUKP, and tested the algorithm on SUKP instances with up to 5,000 items and elements. For the BMCP, Li et al.~\citep{Li2021} proposed the first local search method. Zhou et al.~\citep{Zhou2022} proposed a local search algorithm based on a partial depth-first search tree.

Local search algorithms have obtained excellent results for solving the SUKP and BMCP. However, there are still some disadvantages of the existing local search algorithms for these two problems. The I2PLS~\citep{Wei2019}, KBTS~\citep{Wei2021KBTS}, MSBTS~\citep{Wei2021MSBTS}, and PLTS~\citep{Li2021} algorithms all tend to find the best neighbor solution of the current solution in each iteration. However, their search neighborhood contains lots of low-quality moves, which may reduce the algorithm efficiency. The search region of the ATS-DLA algorithm~\citep{Zhou2021} is small, which may make the algorithm hard to escape from some local optima. The VDLS algorithm~\citep{Zhou2022} has a wide and deep search region. However, VDLS does not allow the current solution worse than the previous one, which may restrict the algorithm's search ability. In summary, these local search methods can not trade-off the efficiency and search region well.

To handle this issue, we propose an efficient and effective local search algorithm, called E2LS, for both the SUKP and BMCP. To the best of our knowledge, this is the first time that an algorithm has been proposed to solve the SUKP and BMCP simultaneously. E2LS uses a random greedy algorithm to generate the initial solution, and an effective local search method to explore the solution space. E2LS restricts the items that can be removed from or added into the current solution to improve the search efficiency. In this way, E2LS can refine the search region by abandoning low-quality candidates. The local search operator in E2LS then traverses the refined search region to find high-quality moves. Thus E2LS can explore the solution space widely and quickly, so as to find high-quality solutions. Moreover, the tabu search method in~\citep{Wei2021MSBTS} is used in E2LS to prevent the algorithm from getting stuck in local optima. Indeed, as we have shown in this work, our proposed E2LS algorithm significantly outperforms the state-of-the-art heuristic algorithms for both the SUKP and BMCP.

The main contributions of this work are as follows.

\begin{itemize}
\item We propose an efficient and effective local search algorithm called E2LS for the SUKP and BMCP. We investigate for the first time an algorithm for solving both of these two problems.
\item E2LS trade-offs the search region and search efficiency well. E2LS restricts the items that can be removed from or added into the current solution, so as to abandon low-quality moves and refine the search region. The proposed operator ADD$^*$ can traverse the refined search region, so as to explore the solution space widely and efficiently.
\item The mechanism that trade-offs the search region and efficiency, as well as the method of traversing the refined search region, could be applied to other combinatorial optimization problems.
\item Extensive experiments demonstrate that E2LS significantly outperforms the state-of-the-art algorithms for both the SUKP and BMCP. In particular, E2LS provides four new best-known solutions for the SUKP, and 27 new best-known solutions for the BMCP.
\end{itemize}


\section{The Proposed E2LS Algorithm}
\label{Sec_Method}
This section introduces the proposed E2LS algorithm. We first present the components of E2LS, including the random greedy initialization method and the searching methods, then present the main process of E2LS. Finally, we discuss the advantages of E2LS over other state-of-the-art local search algorithms for the SUKP and BMCP. Note that the proposed E2LS algorithm can be used to solve both the SUKP and BMCP. This section mainly introduces its SUKP version. The BMCP version can be obtained simply by swapping the optimization objective and constraint objective of the SUKP version.

Before introducing our method, we first present several essential definitions used in the E2LS algorithm. 

\textbf{Definition 1. (Additional Weight)} Note that $W(S)$ (see Eq. \ref{eq_W}) is the total weight of the elements covered by the items in $S$. Let $AW(S,i_j)$ to be the additional weight of an item $i_j$ to a solution $S$. If $i_j \notin S$, the additional weight $AW(S,i_j) = W(S \cup \{i_j\}) - W(S)$ represents the increase of the total weight of the covered elements caused by adding $i_j$ into $S$. Otherwise, $AW(S,i_j) = W(S) - W(S \backslash \{i_j\})$ represents the decrease of the total weight of the covered elements caused by removing $i_j$ from $S$.

\textbf{Definition 2. (Value-weight Ratio)} The value-weight ratio of an item $i_j$ to a solution $S$ is defined as $R_{vw}(S,i_j) = v_j / AW(S,i_j)$, which is the ratio of the value of $i_j$ to the addition weight of $i_j$ to $S$. Obviously, an item with a larger value-weight ratio to a solution $S$ is a better candidate item of $S$~\citep{Khuller1999,Zhou2021,Zhou2022}.


\subsection{Random Greedy Initialization Method}
We propose a simple construction method to generate the initial solution for E2LS. The procedure of the initialization method is shown in Algorithm \ref{alg_init}. 

\begin{algorithm}[t]
\caption{Random\_Greedy($t$)}
\label{alg_init}
\LinesNumbered 
\KwIn{Sampling times $t$}
\KwOut{Solution $S$}
Initialize $S \leftarrow \emptyset$\;
\While{TRUE}{
Initialize feasible candidates $FC \leftarrow \emptyset$\;
\For{$j \leftarrow 1 : n$}{
\lIf{$i_j \in S$}{\textbf{continue}}
\lIf{$AW(S,i_j) = 0$}{$S \leftarrow S \cup \{i_j\}$}
\lIf{$AW(S,i_j) + W(S) \leq C$}{$FC \leftarrow FC \cup \{i_j\}$}
}
\lIf{$FC = \emptyset$}{\textbf{break}}
\Else{
$M \leftarrow 0$\;
\For{$j \leftarrow 1 : t$}{
$i_r \leftarrow$ a random item in $FC$\;
\If{$R_{vw}(S,i_r) > M$}{$M \leftarrow R_{vw}(S,i_r), i_b \leftarrow i_r$}
}
$S \leftarrow S \cup \{i_b\}$\;
}
}
\textbf{return} $S$\;
\end{algorithm}

The algorithm starts with an empty solution $S$, and repeats to add items into $S$ until no more items can be added into $S$ (line 8). In each loop, each item $i_j \notin S$ with $AW(S,i_j) = 0$ will be added into $S$ (line 6). Such an operation can increase $f(S)$ (see Eq. \ref{eq_f}) without increasing $W(S)$. When there is at least one item that adding into $S$ will result in a feasible solution, i.e., the feasible candidates $FC \neq \emptyset$, the algorithm applies the probabilistic sampling strategy~\citep{Cai2015,Zheng2021} to select the item to be added. Specifically, the algorithm first random samples $t$ items with replacement in $FC$ (lines 11-12), then selects to add the item with the maximum value-weight ratio (lines 13-15).

We set the parameter $t$ to be $\sqrt{max\{m,n\}}$ as the algorithm MSBTS~\citep{Wei2021MSBTS} does. This setting can help the E2LS algorithm yield high-quality and diverse initial solutions. In particular, we analyze the influence of $t$ on the performance of E2LS in experiments. The results show that E2LS is very robust and not sensitive to the parameter $t$. Even with a random initialization method (i.e., $t = 1$), E2LS also has excellent performance.

\subsection{Searching Methods in E2LS}
The local search method in E2LS is the main improvement of E2LS to other local search algorithms for the SUKP and BMCP. In E2LS, we propose an efficient and effective local search operator to explore the solution space. We also apply the solution-based tabu search method in~\citep{Wei2021MSBTS} to avoid getting stuck in local optima. This subsection first describes how to represent the tabu list, then introduces the local search process.

\subsubsection{Tabu List Representation}
The tabu search method in~\citep{Wei2021MSBTS} uses three hash vectors $H_1,H_2,H_3$ to represent the tabu list. The length of each vector is set to $L$ ($L = 10^8$ by default). The three hash vectors are initialized to 0, indicating that no solution is prohibited by the tabu list. Each solution $S$ is corresponding to three hash values $h_1(S),h_2(S),h_3(S)$. A solution $S$ is prohibited (i.e., in the tabu list) if $H_1[h_1(S)] \wedge H_2[h_2(S)] \wedge H_3[h_3(S)] = 1$. See the details for calculating the hash values of a solution below.

For an instance with $m$ items, a weight matrix $\mathcal{W} \in \mathbf{N}^{\{3 \times m\}}$ is calculated as follows. First, let $\mathcal{W}_{lj} = \lfloor j^{\gamma_l} \rfloor, l = (1,2,3), j = (1,...,m)$, where $\gamma_1,\gamma_2,\gamma_3$ are set to 1.2, 1.6, 2.0, respectively. Then, random shuffle each of the three rows of $\mathcal{W}$. Given a solution $S$ that can be represented by $(y_1,...,y_m)$ where $y_j = 1$ if $i_j \in S$, and $y_j = 0$ otherwise. The hash values $h_l(S), l = (1,2,3)$ can be calculated by $h_l(S) = (\sum_{j=1}^{m}{\lfloor \mathcal{W}_{lj} \times y_j \rfloor) ~\text{mod}~ L}$.


\subsubsection{Local Search Process}
The procedure of the local search method in E2LS is shown in Algorithm \ref{alg_LS}. The search operator first removes an item from the input solution $S$ (line 6), and then uses the proposed operator ADD$^*$ (Algorithm \ref{alg_add}) to add items into the resulting solution $S'$ (line 8). The best solution that is not in the tabu list found during the search process is represented by $S_b$ (line 9). 

\begin{algorithm}[t]
\caption{Local\_Search($S,r_{num},a_{num}$)}
\label{alg_LS}
\LinesNumbered 
\KwIn{Input solution $S$, maximum size of the candidate set of the items to be removed $r_{num}$, maximum size of the candidate set of the items to be added $a_{num}$}
\KwOut{Output solution $S$}
$S_b \leftarrow \emptyset$\;
$U \leftarrow$ a set of items in $S$\;
Sort the items in $U$ in ascending order of their value-weight ratios to $S$\;
\For{$j \leftarrow 1 : \text{min}\{r_{num},|U|\}$}{
$i \leftarrow$ the $j$-th item in $U$\;
$S' \leftarrow S \backslash \{i\}$\;
\If{$S'$ is in the tabu list}{
$S' \leftarrow$ ADD$^*$($S',\emptyset,a_{num}$)\;
\lIf{$f(S') > f(S_b)$}{$S_b \leftarrow S'$}
}
}

\textbf{return} $S_b$\;
\end{algorithm}

\begin{algorithm}[t]
\caption{ADD$^*$($S,S_b,a_{num}$)}
\label{alg_add}
\LinesNumbered 
\KwIn{Input solution $S$, best solution found in this step $S_b$, maximum size of the candidate set of the items to be added $a_{num}$}
\KwOut{Output solution $S$}
\While{TRUE}{
Initialize feasible candidates $FC \leftarrow \emptyset$\;
\For{$j \leftarrow 1 : m$}{
\lIf{$i_j \in S$}{\textbf{continue}}
\If{$AW(S,i_j) = 0$}{
\If{$S \cup \{i_j\}$ is not in the tabu list}{
$S \leftarrow S \cup \{i_j\}$\;
\lIf{$f(S) > f(S_b)$}{$S_b \leftarrow S$}
}
}
\ElseIf{$AW(S,i_j) + W(S) \leq C$}{$FC \leftarrow FC \cup \{i_j\}$}
}
\lIf{$FC = \emptyset$}{\textbf{return} $S_b$}
}
Sort the items in $FC$ in descending order of their value-weight ratios to $S$\;
\For{$j \leftarrow 1 : \text{min}\{a_{num},|FC|\}$}{
$i \leftarrow$ the $j$-th item in $FC$\;
\lIf{$S \cup \{i\}$ is in the tabu list}{\textbf{continue}}
$S' \leftarrow S \cup \{i\}$\;
\lIf{$f(S') > f(S_b)$}{$S_b \leftarrow S'$}
$S' \leftarrow$ ADD$^*$($S',S_b,a_{num}$)\;
}
\textbf{return} $S_b$\;
\end{algorithm}

As shown in Algorithm \ref{alg_add}, the function ADD$^*$ tries to add items into the input solution $S$ recursively (line 18) until no more items can be added into $S$ (line 11). The best solution that is not in the tabu list found during the process is represented by $S_b$ (lines 8 and 17). The algorithm first calculates the set of the feasible candidate items $FC$ of $S$ (lines 1-11). During the process, each item $i_j \notin S$ with $AW(S,i_j) = 0$ will be added into $S$ if the resulting solution is not in the tabu list (lines 5-7). After that, the algorithm traverses part of the set $FC$ to add an item $i$ into $S$ (lines 13-14), and then calls the function ADD$^*$ with the input solution $S \cup \{i\}$ to continue the search (line 18). During the process, the solutions in the tabu list are prohibited (line 15).

The proposed function ADD$^*$ is powerful and actually has a wide search region. The reasons are as follows. Suppose we set $a_{num} = m$, ADD$^*$ can traverse all the combinations of the candidate items by the recursive process, i.e., ADD$^*$ can find the optimal solution corresponding to its input partial solution when the items in its input solution are fixed. In particular, the function ADD$^*$ can be regarded as an exact solver for an instance with $m$ items if we call ADD$^*$($\emptyset,\emptyset,m$). However, an exhaustive search must be inefficient.

In order to improve the efficiency of the local search process, the low-quality moves (candidate items) should not be considered during the search process. To yield high-quality results, the items with large value-weight ratios to the current solution should be added into (or be kept in) it, and the items with small value-weight ratios should be removed from (or not be added into) it. Therefore, in Algorithm \ref{alg_LS}, the items that can be removed from $S$ are restricted by a parameter $r_{num}$. That is, only the top min$\{r_{num},|S|\}$ items with minimum value-weight ratios to $S$ can be removed from $S$ (lines 3-4). In Algorithm \ref{alg_add}, only the top min$\{a_{num},|FC|\}$ items with maximum value-weight ratios to $S$ can be added into $S$ (lines 12-13). By applying this strategy, the search region can be refined significantly, i.e., the number of items considered by ADD$^*$ is reduced from $|FC|$ to min$\{a_{num},|FC|\}$, and the efficiency can be improved greatly. Also, the operator ADD$^*$ does not traverse all the moves, but only traverses the refined search region.

Moreover, our proposed ADD$^*$ operator can add multiple items into the current solution $S$, which can not be regarded as the same as the combination of multiple continuous $ADD$ operators in~\citep{Wei2019,Wei2021KBTS} or $flip$ operators (on items not in $S$) in ~\citep{Wei2021MSBTS,Li2021}. For example, the ADD$^*$ operator adds items $i_1,i_2$, and $i_3$ into $S$, while the best neighbor solution of $S$ find by the commonly used $ADD$ operator might chooses to add item $i_4$. In summary, ADD$^*$ effective because it can traverse various combinations of the items in the refined search region that can be added into the current solution.

\subsubsection{Main Framework of E2LS}
The main process of E2LS is shown in Algorithm \ref{alg_E2LS}. E2LS first initializes the three hash vectors $H_1,H_2,H_3$ to 0 (line 2), and calls the Random\_Greedy function (Algorithm \ref{alg_init}) to generate the initial solution $S$ (line 3). Then, E2LS calls the Local\_Search function (Algorithm \ref{alg_LS}) to explore the solution space until the cut-off time is reached (lines 5-12). During the iterations, the Random\_Greedy function will be called again when the Local\_Search function with the input solution $S$ can not find a solution that is not in the tabu list (line 7-8), i.e., the output solution of the Local\_Search function is $\emptyset$. The solution $S$ obtained in each iteration will be added into the tabu list by setting $H_1[h_1(S)] \wedge H_2[h_2(S)] \wedge H_3[h_3(S)]$ to 1 (line 11). The best solution found so far is represented by $S_b$ (line 12).

\begin{algorithm}[t]
\caption{E2LS($I,T_{max},t,r_{num},a_{num}$)}
\label{alg_E2LS}
\LinesNumbered 
\KwIn{Instance $I$, cut-off time $T_{max}$, sampling times $t$, maximum size of the candidate set of the items to be removed $r_{num}$, maximum size of the candidate set of the items to be added $a_{num}$}
\KwOut{Output solution $S$}
Read instance $I$\;
Initialize $H_1,H_2,H_3$ to 0\;
$S \leftarrow$ Random\_Greedy($t$)\;
Initialize $S_b \leftarrow \emptyset$\;
\While{the cut-off time $T_{max}$ is not reached}{
$S' \leftarrow$ Local\_Search($S,r_{num},a_{num}$)\;
\If{$S' = \emptyset$}{
$S \leftarrow$ Random\_Greedy($t$)\;
\textbf{continue}\;
}
$S \leftarrow S'$\;
\lFor{$j \leftarrow 1 : 3$}{$H_j[h_j(S)] \leftarrow 1$}
\lIf{$f(S) > f(S_b)$}{$S_b \leftarrow S$}
}
\textbf{return} $S_b$\;
\end{algorithm}

\subsubsection{Advantages of E2LS}
This subsection discusses the main advantages of our proposed E2LS over the state-of-the-art local search algorithms for the SUKP and BMCP.

The first category of local search methods includes the I2PLS~\citep{Wei2019}, KBTS~\citep{Wei2021KBTS}, MSBTS~\citep{Wei2021MSBTS} algorithms for the SUKP, and the PLTS~\citep{Li2021} algorithm for the BMCP. These algorithms are all based on tabu search. Their common shortcoming is that their search neighborhoods contain lots of low-quality moves. While E2LS refines the search neighborhood according to the value-weight ratios of the items. Thus, E2LS shows much better performance and higher efficiency than these algorithms.

The second category is the ATS-DLA algorithm~\citep{Zhou2021}, which also uses the tabu search method and is efficient for large scale SUKP instances. However, ATS-DLA can only remove or add one item in each iteration, which leads to a relatively small and overrefined search region, and may make it hard to escape from local optima in some cases. While E2LS can add multiple items in each iteration by traversing the refined search region, and considering various combinations of the items. Therefore, E2LS can explore the solution space wider and deeper than ATS-DLA, so as to find higher-quality solutions. 

The third category is the VDLS algorithm~\citep{Zhou2022}, which is not based on tabu search but a partial depth-first search method. VDLS has a large search region, and can yield significantly better solutions than the PLTS algorithm~\citep{Li2021}. However, VDLS does not allow the solution to get worse during the search process, and the initial solution generated by VDLS is fixed. Such mechanisms limit the search ability of the algorithm. E2LS does not restrict that the current solution must be better than the previous one, and applies the tabu search method to avoid getting stuck in local optima. Moreover, the random greedy initialization method in E2LS can generate high-quality and diverse initial solutions for E2LS. Thus E2LS also shows significantly better performance than VDLS.




\section{Computational Results}
\label{Sec_Result}
This section presents experimental results and analyses, before which we first introduce the benchmark instances used in the experiments and the experimental setup.

\subsection{Benchmark Instances}
We tested E2LS on a total of 78 public SUKP instances with at most 5,000 items or elements, and 90 public BMCP instances with no more than 5,200 items or elements. The 78 tested SUKP instances can be divided into three sets as follows.

\begin{itemize}
\item \textit{Set I:} This set contains 30 instances with 85 to 500 items or elements. This set was proposed in~\citep{He2018} and widely used in~\citep{He2018,Ozsoydan2019,Lin2019,Wei2019,Dahmani2020,Wu2020,Wei2021KBTS,Wei2021MSBTS}.
\item \textit{Set II:} This set contains 30 instances with 585 to 1,000 items or elements. This set was proposed in~\citep{Wei2021KBTS} and used in~\citep{Wei2021KBTS,Wei2021MSBTS}.
\item \textit{Set III:} This set contains 18 instances with 850 to 5,000 items or elements. This set was proposed in~\citep{Zhou2021}.
\end{itemize}

Each instance in \textit{Sets I}, \textit{II}, and \textit{III} is characterized by four parameters: the number of items $m$, the number of elements $n$, the density of the relation matrix $\alpha = (\sum_{j=1}^m{\sum_{k=1}^n{R_{ij}}})/(mn)$, and the ratio of knapsack capacity $C$ to the total weight of the elements $\beta = C/\sum_{k=1}^n{w_k}$. The name of a SUKP instance consists of these four parameters. For example, \textit{sukp\_85\_100\_0.10\_0.75} represents a SUKP instance with 85 items, 100 elements, $\alpha = 0.10$, and $\beta = 0.75$. 

The 90 tested BMCP instances can be divided into three sets as follows.

\begin{itemize}
\item \textit{Set A:} This set contains 30 instances with 585 to 1,000 items or elements. This set was proposed in~\citep{Li2021} and used in~\citep{Li2021,Zhou2022}.
\item \textit{Set B:} This set contains 30 instances with the number of items or elements ranging from 1,000 to 1,600. This set was proposed in~\citep{Zhou2022}.
\item \textit{Set C:} This set contains 30 instances with the number of items or elements ranging from 4,000 to 5,200. This set was proposed in~\citep{Zhou2022}.
\end{itemize}

Each instance in \textit{Set A} is characterized by four parameters: the number of items $m$, the number of elements $n$, the knapsack capacity (budget) $C$, and the density of the relation matrix $\alpha$. The instances in \textit{Sets B} and \textit{C} have more complex structures than those in \textit{Set A}. Zhou et al.~\citep{Zhou2022} created these instances by first randomly grouping the items and elements, then deciding the connection of them according to a parameter $\rho$ that represents the density of the relation matrix of each group. Each instance in \textit{Sets B} and \textit{C} is characterized by parameters $m,n,C$ and $\rho$.

\subsection{Experimental Setup}
We first introduce the baseline algorithms we selected. For the SUKP, we select some of the state-of-the-art heuristic algorithms, including ATS-DLA~\citep{Zhou2021}, MSBTS~\citep{Wei2021MSBTS}, KBTS~\citep{Wei2021KBTS}, and I2PLS~\citep{Wei2019}, as the baseline algorithms. For the BMCP, we select the heuristic algorithms PLTS~\citep{Li2021} and VDLS~\citep{Zhou2022} as the baseline algorithms. To our knowledge, they are also the only two heuristic algorithms for the BMCP.

The E2LS algorithm and the baseline algorithms were implemented in C++ and compiled by g++. All the experiments were performed on a server using an Intel® Xeon® E5-1603 v3 2.80GHz CPU, running Ubuntu 16.04 Linux operation system. The parameters in E2LS include the sampling times $t$, and the maximum size of the candidate set of the items to be removed/added $r_{num}/a_{num}$. We tuned these parameters according to our experience. The default value of $t$ is set to $\sqrt{max\{m,n\}}$. For the parameters $r_{num}$ and $a_{num}$, we set the default values of them to be different when solving the SUKP and BMCP, since the properties of these two problems are different. Specifically, for the SUKP, we set the default values as $r_{num} = 2$ and $a_{num} = 2$. For the BMCP, we set $r_{num} = 5$ and $a_{num} = 5$. The detailed reasons why their default values are different when solving the SUKP and BMCP, as well as the influence of these parameters on the performance of E2LS, are presented in Section \ref{Sec_para}.

We set the cut-off time for each algorithm to be 500 seconds for the instances in \textit{Set I} as~\citep{Wei2021MSBTS} did, 1,000 seconds for the instances in \textit{Sets II} and \textit{III} as~\citep{Wei2021MSBTS,Zhou2021} did, 600 seconds for the instances in \textit{Set A} as~\citep{Li2021} did, and 1,800 seconds for the instances in \textit{Sets B} and \textit{C} as~\citep{Zhou2022} did. Each instance was calculated 10 independent times for each algorithm. 

\begin{sidewaystable}[thp]
\caption{Comparison of E2LS with the baseline algorithms on 30 SUKP instances of \textit{Set I}. Unique best results appear in bold. Equal best results appear in italic.\vspace{-0.3em}}
\label{table-SUKP1}
\centering
\scalebox{0.55}{
\begin{tabular}{lrrrrrrrrrrrrrrrrrrrrrrrrr} \bottomrule
Instance\textbackslash{}Algorithm &  & \multicolumn{4}{r}{I2PLS}                             &  & \multicolumn{4}{r}{KBTS}                                &  & \multicolumn{4}{r}{MSBTS}                             &  & \multicolumn{4}{r}{ATS-DLA}                           &  & \multicolumn{4}{r}{E2LS}                                \\ \cline{3-6} \cline{8-11} \cline{13-16} \cline{18-21} \cline{23-26} 
                                  &  & Best           & Average           & S.D.   & Time    &  & Best              & Average           & S.D.  & Time    &  & Best           & Average           & S.D.   & Time    &  & Best           & Average           & S.D.   & Time    &  & Best              & Average           & S.D.  & Time    \\ \hline
sukp\_100\_85\_0.10\_0.75.txt     &  & \textit{13283} & \textit{13283.00} & 0.00   & 0.883   &  & \textit{13283}    & \textit{13283.00} & 0.00  & 3.219   &  & \textit{13283} & \textit{13283.00} & 0.00   & 497.286 &  & 13016          & 12917.50          & 101.70 & 5.477   &  & \textit{13283}    & 13276.60          & 12.80 & 66.663  \\
sukp\_100\_85\_0.15\_0.85.txt     &  & \textit{12479} & 12361.10          & 128.09 & 42.547  &  & \textit{12479}    & \textbf{12479.00} & 0.00  & 22.052  &  & \textit{12479} & 12445.40          & 69.21  & 498.053 &  & 11902          & 11400.20          & 369.75 & 0.013   &  & \textit{12479}    & 12456.70          & 55.14 & 134.929 \\
sukp\_200\_185\_0.10\_0.75.txt    &  & \textit{13521} & \textit{13521.00} & 0.00   & 50.447  &  & \textit{13521}    & \textit{13521.00} & 0.00  & 0.884   &  & \textit{13521} & \textit{13521.00} & 0.00   & 406.291 &  & \textit{13521} & \textit{13521.00} & 0.00   & 51.034  &  & \textit{13521}    & \textit{13521.00} & 0.00  & 17.815  \\
\end{tabular}}
\end{sidewaystable}

\begin{sidewaystable}[thp]
\caption{Comparison of E2LS with the baseline algorithms on 30 SUKP instances of \textit{Set II}. Unique best results appear in bold. Equal best results appear in italic.\vspace{-0.3em}}
\label{table-SUKP2}
\centering
\scalebox{0.55}{\begin{tabular}{lrrrrrrrrrrrrrrrrrrrrrrrrr} \bottomrule
Instance\textbackslash{}Algorithm &  & \multicolumn{4}{r}{I2PLS}                            &  & \multicolumn{4}{r}{KBTS}                             &  & \multicolumn{4}{r}{MSBTS}                            &  & \multicolumn{4}{r}{ATS-DLA}                          &  & \multicolumn{4}{r}{E2LS}                              \\ \cline{3-6} \cline{8-11} \cline{13-16} \cline{18-21} \cline{23-26} 
                                  &  & Best           & Average          & S.D.   & Time    &  & Best           & Average           & S.D.  & Time    &  & Best           & Average           & S.D.  & Time    &  & Best           & Average           & S.D.  & Time    &  & Best             & Average           & S.D. & Time    \\ \hline
sukp\_600\_585\_0.10\_0.75.txt    &  & 9750           & 9731.50          & 12.64  & 315.869 &  & \textit{9914}  & \textit{9914.00}  & 0.00  & 183.513 &  & \textit{9914}  & \textit{9914.00}  & 0.00  & 152.437 &  & 9790           & 9746.70           & 14.46 & 269.974 &  & \textit{9914}    & \textit{9914.00}  & 0.00 & 16.191  \\
sukp\_600\_585\_0.15\_0.85.txt    &  & \textit{9357}  & 9320.00          & 14.57  & 189.829 &  & \textit{9357}  & \textit{9357.00}  & 0.00  & 370.468 &  & \textit{9357}  & \textit{9357.00}  & 0.00  & 49.735  &  & \textit{9357}  & \textit{9357.00}  & 0.00  & 4.552   &  & \textit{9357}    & \textit{9357.00}  & 0.00 & 27.229  \\
sukp\_700\_685\_0.10\_0.75.txt    &  & \textit{9881}  & 9814.80          & 43.61  & 328.507 &  & \textit{9881}  & 9853.00           & 18.33 & 244.831 &  & \textit{9881}  & \textit{9881.00}  & 0.00  & 22.714  &  & \textit{9881}  & \textit{9881.00}  & 0.00  & 1.810   &  & \textit{9881}    & \textit{9881.00}  & 0.00 & 16.557  \\
\end{tabular}}
\end{sidewaystable}

\begin{sidewaystable}[thp]
\caption{Comparison of E2LS with the baseline algorithms on 18 SUKP instances of \textit{Set III}. Unique best results appear in bold. Equal best results appear in italic.\vspace{-0.3em}}
\label{table-SUKP3}
\centering
\scalebox{0.6}{\begin{tabular}{lrrrrrrrrrrrrrrrrrrrrrrrrr} \bottomrule
Instance\textbackslash{}Algorithm &  & \multicolumn{4}{r}{I2PLS}             &  & \multicolumn{4}{r}{KBTS}                             &  & \multicolumn{4}{r}{MSBTS}                             &  & \multicolumn{4}{r}{ATS-DLA}                          &  & \multicolumn{4}{r}{E2LS}                               \\ \cline{3-6} \cline{8-11} \cline{13-16} \cline{18-21} \cline{23-26} 
                                  &  & Best    & Average  & S.D.   & Time    &  & Best           & Average          & S.D.   & Time    &  & Best           & Average           & S.D.   & Time    &  & Best           & Average           & S.D.  & Time    &  & Best             & Average           & S.D.  & Time    \\ \hline
sukp\_850\_1000\_0.10\_0.75.txt   &  & 9944    & 9851.90  & 96.12  & 454.059 &  & 9983           & 9942.20          & 21.77  & 183.600 &  & \textit{10028} & \textit{10028.00} & 0.00   & 189.841 &  & \textit{10028} & \textit{10028.00} & 0.00  & 16.553  &  & \textit{10028}   & \textit{10028.00} & 0.00  & 85.080  \\
sukp\_850\_1000\_0.15\_0.85.txt   &  & 8851    & 8838.30  & 22.52  & 355.332 &  & \textit{8957}  & 8919.10          & 17.82  & 173.008 &  & \textit{8957}  & \textit{8957.00}  & 0.00   & 356.205 &  & \textit{8957}  & \textit{8957.00}  & 0.00  & 17.191  &  & \textit{8957}    & \textit{8957.00}  & 0.00  & 39.932  \\
sukp\_1000\_850\_0.10\_0.75.txt   &  & 10464   & 10434.30 & 46.83  & 344.512 &  & \textit{10623} & 10610.40         & 10.29  & 309.152 &  & \textit{10623} & \textit{10623.00} & 0.00   & 385.623 &  & 10511          & 10403.80          & 39.10 & 343.725 &  & \textit{10623}   & \textit{10623.00} & 0.00  & 51.127  \\
\end{tabular}}
\end{sidewaystable}
\begin{table*}[!t]
\caption{Comparison of E2LS with the baseline algorithms on 30 BMCP instances of \textit{Set A}. Unique/Equal best results appear in bold/italic.}
\label{table-BMCP1}
\centering
\scalebox{0.65}{\begin{tabular}{lrrrrrrrrrrr} \bottomrule
Instance\textbackslash{}Algorithm &  & \multicolumn{2}{r}{PLTS}   &  & \multicolumn{2}{r}{VDLS}              &           & \multicolumn{4}{r}{E2LS}                               \\ \cline{3-4} \cline{6-7} \cline{9-12} 
                                  &  & Best           & Average   &  & Best              & Average           &           & Best              & Average            & S.D. & Time   \\ \hline
bmcp\_585\_600\_0.05\_2000        &  & 70997          & 70892.60  &  & \textit{71102}    & 71034.37          &           & \textit{71102}    & \textbf{71102.00}  & 0.00 & 0.525  \\
bmcp\_585\_600\_0.075\_1500       &  & 70677          & 70535.90  &  & \textit{71025}    & 70492.53          &           & \textit{71025}    & \textbf{71025.00}  & 0.00 & 0.729  \\
bmcp\_600\_585\_0.05\_2000        &  & 67187          & 66807.10  &  & \textit{67636}    & 67621.60          &           & \textit{67636}    & \textbf{67636.00}  & 0.00 & 0.317  \\
bmcp\_600\_585\_0.075\_1500       &  & 70318          & 70318.00  &  & \textit{70588}    & 70277.00          &           & \textit{70588}    & \textbf{70588.00}  & 0.00 & 2.357  \\
bmcp\_600\_600\_0.05\_2000        &  & 68453          & 68346.60  &  & \textit{68738}    & 68580.70          &           & \textit{68738}    & \textbf{68738.00}  & 0.00 & 0.419  \\
bmcp\_600\_600\_0.075\_1500       &  & 71746          & 71746.00  &  & \textit{71904}    & 71780.73          &           & \textit{71904}    & \textbf{71904.00}  & 0.00 & 0.476  \\
\end{tabular}}
\end{table*}

\begin{table*}[!t]
\caption{Comparison of E2LS with the baseline algorithms on 30 BMCP instances of \textit{Set B}. Unique/Equal best results appear in bold/italic.}
\label{table-BMCP2}
\centering
\scalebox{0.65}{\begin{tabular}{lrrrrrrrrrrr} \bottomrule
Instance\textbackslash{}Algorithm &  & \multicolumn{2}{r}{PLTS}    &  & \multicolumn{2}{r}{VDLS}             &           & \multicolumn{4}{r}{E2LS}                                 \\ \cline{3-4} \cline{6-7} \cline{9-12} 
                                  &  & Best            & Average   &  & Best            & Average            &           & Best               & Average            & S.D. & Time    \\ \hline
bmcp\_1100\_1000\_0.3\_3000       &  & 142835          & 142428.70 &  & \textit{143475} & 143224.60          &           & \textit{143475}    & \textbf{143475.00} & 0.00 & 1.111   \\
bmcp\_1100\_1000\_0.5\_2000       &  & 125974          & 125790.00 &  & 126917          & 126021.57          &           & \textbf{127157}    & \textbf{127157.00} & 0.00 & 0.839   \\
bmcp\_1100\_1100\_0.3\_3000       &  & 154936          & 154661.70 &  & \textit{157530} & 156462.53          &           & \textit{157530}    & \textbf{157530.00} & 0.00 & 1.006   \\
bmcp\_1100\_1100\_0.5\_2000       &  & 139196          & 138169.00 &  & 139141          & 138719.40          &           & \textbf{139235}    & \textbf{139235.00} & 0.00 & 1.813   \\
bmcp\_1100\_1200\_0.3\_3000       &  & 169113          & 168763.60 &  & \textit{172660} & 172075.30          &           & \textit{172660}    & \textbf{172660.00} & 0.00 & 0.442   \\
bmcp\_1100\_1200\_0.5\_2000       &  & 150301          & 149366.40 &  & \textit{150493} & 149944.03          &           & \textit{150493}    & \textbf{150493.00} & 0.00 & 0.636   \\
\end{tabular}}
\end{table*}

\begin{table*}[!t]
\caption{Comparison of E2LS with the baseline algorithms on 30 BMCP instances of \textit{Set C}. Unique/Equal best results appear in bold/italic.}
\label{table-BMCP3}
\centering
\scalebox{0.65}{\begin{tabular}{lrrrrrrrrrrr} \bottomrule
Instance\textbackslash{}Algorithm &  & \multicolumn{2}{r}{PLTS} &  & \multicolumn{2}{r}{VDLS}      &  & \multicolumn{4}{r}{E2LS}                                    \\ \cline{3-4} \cline{6-7} \cline{9-12} 
                                  &  & Best        & Average    &  & Best             & Average    &  & Best                & Average             & S.D.  & Time    \\ \hline
bmcp\_4200\_4000\_0.3\_10000      &  & 1778013     & 1747411.90 &  & \textit{1835947} & 1833108.20 &  & \textit{1835947}    & \textbf{1835947.00} & 0.00  & 5.636   \\
bmcp\_4200\_4000\_0.5\_7000       &  & 1685141     & 1666673.30 &  & \textit{1738553} & 1729188.00 &  & \textit{1738553}    & \textbf{1738553.00} & 0.00  & 17.241  \\
bmcp\_4200\_4200\_0.3\_10000      &  & 1874190     & 1839780.10 &  & 1925706          & 1918624.23 &  & \textbf{1925859}    & \textbf{1925859.00} & 0.00  & 3.422   \\
bmcp\_4200\_4200\_0.5\_7000       &  & 1764277     & 1742121.70 &  & 1811649          & 1801880.20 &  & \textbf{1816714}    & \textbf{1816714.00} & 0.00  & 23.069  \\
bmcp\_4200\_4400\_0.3\_10000      &  & 1954069     & 1932251.00 &  & \textit{2023093} & 2020209.80 &  & \textit{2023093}    & \textbf{2023093.00} & 0.00  & 21.845  \\
bmcp\_4200\_4400\_0.5\_7000       &  & 1853677     & 1820444.40 &  & \textit{1903515} & 1889809.53 &  & \textit{1903515}    & \textbf{1903515.00} & 0.00  & 17.612  \\
\end{tabular}}
\end{table*}

\begin{table}[t]
\caption{Summarized comparisons of E2LS against each baseline algorithm over the three sets of SUKP benchmark instances.}
\label{table-SUKP-compare}
\centering
\scalebox{0.7}{\begin{tabular}{lllrrr} \bottomrule
Algorithm pair   & Benchmark        & Indicator        & \#Wins & \#Ties & \#Losses \\ \hline
E2LS vs. I2PLS   & \textit{Set I}   & Best solution    & 3      & 27     & 0        \\
                 & \textit{}        & Average solution & 14     & 14     & 2        \\
                 & \textit{Set II}  & Best solution    & 18     & 12     & 0        \\
                 & \textit{}        & Average solution & 29     & 1      & 0        \\
                 & \textit{Set III} & Best solution    & 18     & 0      & 0        \\
                 & \textit{}        & Average solution & 18     & 0      & 0        \\ \hline
E2LS vs. KBTS    & \textit{Set I}   & Best solution    & 0      & 30     & 0        \\
\end{tabular}}
\end{table}
\begin{table}[t]
\caption{Summarized comparisons of E2LS against each baseline algorithm over the three sets of BMCP benchmark instances.}
\label{table-BMCP-compare}
\centering
\scalebox{0.7}{\begin{tabular}{lllrrr} \bottomrule
Algorithm pair & Benchmark      & Indicator        & \#Wins & \#Ties & \#Losses \\ \hline
E2LS vs. PLTS  & \textit{Set A} & Best solution    & 28     & 2      & 0        \\
               & \textit{}      & Average solution & 30     & 0      & 0        \\
               & \textit{Set B} & Best solution    & 28     & 2      & 0        \\
               & \textit{}      & Average solution & 30     & 0      & 0        \\
               & \textit{Set C} & Best solution    & 30     & 0      & 0        \\
               & \textit{}      & Average solution & 30     & 0      & 0        \\ \hline
E2LS vs. VDLS  & \textit{Set A} & Best solution    & 0      & 30     & 0        \\
               & \textit{}      & Average solution & 28     & 2      & 0        \\
               & \textit{Set B} & Best solution    & 7      & 23     & 0        \\
               & \textit{}      & Average solution & 29     & 1      & 0        \\
               & \textit{Set C} & Best solution    & 20     & 10      & 0        \\
               & \textit{}      & Average solution & 30     & 0      & 0       \\ \toprule
\end{tabular}}
\end{table}

\subsection{Comparison with the baselines}
The comparison results of E2LS with the baseline algorithms on the three sets of SUKP instances are presented in Tables \ref{table-SUKP1}, \ref{table-SUKP2}, and \ref{table-SUKP3}\footnote{Note that there are two instances with the same name but the different structures in \textit{Set II} and \textit{Set III}.}, respectively. The comparison results of E2LS with the baseline algorithms on the three sets of BMCP instances are presented in Tables \ref{table-BMCP1}, \ref{table-BMCP2}, and \ref{table-BMCP3}, respectively. In these tables, unique best results are appeared in bold, while equal best results are appeared in italic. Tables \ref{table-SUKP1}, \ref{table-SUKP2}, and \ref{table-SUKP3} compare the best solution (objective value), average solution, standard deviations over 10 runs (S.D.), and the average run times in seconds (to obtain the best solution in each run) of each involved algorithm. Tables \ref{table-BMCP1}, \ref{table-BMCP2}, and \ref{table-BMCP3} present the best solution and average solution of each involved algorithm, coupled with the standard deviations and average run times of E2LS. The results of PLTS and VDLS in Tables \ref{table-BMCP1}, \ref{table-BMCP2}, and \ref{table-BMCP3} are from~\citep{Zhou2022} (they used the similar machine as ours, Intel® Xeon® E5-2650 v3 2.30GHz).

Moreover, in order to show the advantage of E2LS over the baselines more clearly, we summarize the comparative results between E2LS and each baseline algorithm in Tables \ref{table-SUKP-compare} and \ref{table-BMCP-compare}. Columns \#Wins, \#Ties, and \#Losses indicate the number of instances for which E2LS obtains a better, equal, and worse result than the compared algorithm according to the best solution and average solution indicators.

As the results shown in Tables \ref{table-SUKP1}, \ref{table-SUKP2}, \ref{table-SUKP3}, and \ref{table-SUKP-compare}, E2LS does not lose to any baseline algorithm according to the best solution indicator. Actually, E2LS can obtain the best-known solution for all the 78 tested SUKP instances. E2LS also has good stability and robustness, since the average solutions obtained by E2LS are also excellent and the standard deviations of E2LS are very small. In particular, E2LS obtains three new best-known solutions for instances \textit{sukp\_1000\_850\_0.15\_0.85}, \textit{sukp\_3000\_2850\_0.10\_0.75}, and \textit{sukp\_3000\_2850\_0.15\_0.85}\footnote{The result 9565 of instance \textit{sukp\_1000\_850\_0.15\_0.85} can also be obtained by MSBTS, but has not been reported in the literature. The result 9207 of instance \textit{sukp\_5000\_4850\_0.10\_0.75} has been reported in~\citep{Zhou2021}.}.

Moreover, there are no baseline algorithms that can solve the SUKP instances with various scales well. For example, ATS-DLA shows good performance for the instances in \textit{Sets II} and \textit{III}, since its search operator that considers only one item per step makes it efficient for (relatively) large instances. However, ATS-DLA is not good at solving the instances in \textit{Set I}, because its search operator has a small search region, while a large search region is more important than a high search efficiency for small instances. Algorithms KBTS and MSBTS show good performance for the instances in \textit{Set I}, because they traverse all the possible moves per step and thus have large search regions. However, KBTS shows worse performance than MSBTS on \textit{Set II}, since the solution-based tabu search is better than the attributed-based tabu search for the SUKP~\citep{Wei2021MSBTS}. Also, both KBTS and MSBTS show worse performance than ATS-DLA on \textit{Set III}, since their operators that traverse all the possible moves are inefficient for large instances.

While the proposed E2LS shows excellent performance on all three sets of SUKP instances, because E2LS extracts the advantages of the baseline algorithms and improves their disadvantages. On the one hand, E2LS refines the search region according to the value-weight ratios of the items, so as to abandon the low-quality moves and improve the search efficiency. Thus E2LS works well on \textit{Sets II} and \textit{III}. On the other hand, the proposed operator ADD$^*$ can traverse the refined search region and add the best feasible combination of multiple candidate items into the current solution per iteration, which leads to a large search region. Thus E2LS also works well on \textit{Set I}. In summary, E2LS is efficient and effective because it can trade-off the search region and search efficiency well.

As for the comparison results on the BMCP instances shown in Tables \ref{table-BMCP1}, \ref{table-BMCP2}, \ref{table-BMCP3}, and \ref{table-BMCP-compare}, the advantage of E2LS over the BMCP baseline algorithms PLTS and VDLS is more obvious than that of E2LS over the SUKP baselines. This might be because the SUKP is more well-studied than the BMCP in terms of heuristic methods. Specifically, E2LS does not lose to the BMCP baselines according to either the best or the average solution indicator, and obtains 7/20 new best-known solutions for the instances in \textit{B}/\textit{C}. Moreover, the average solutions of E2LS are equal to its best solutions on all the tested instances except \textit{bmcp\_5000\_4800\_0.5\_7000}. And E2LS can obtain such excellent results within very small run times for most of the tested instances. The results indicate again the excellent stability, robustness, and efficiency of the proposed E2LS algorithm.

\begin{figure*}[t]
\centering
\subfigure[Comparison results on \textit{Set I}]{
\includegraphics[width=0.96\columnwidth]{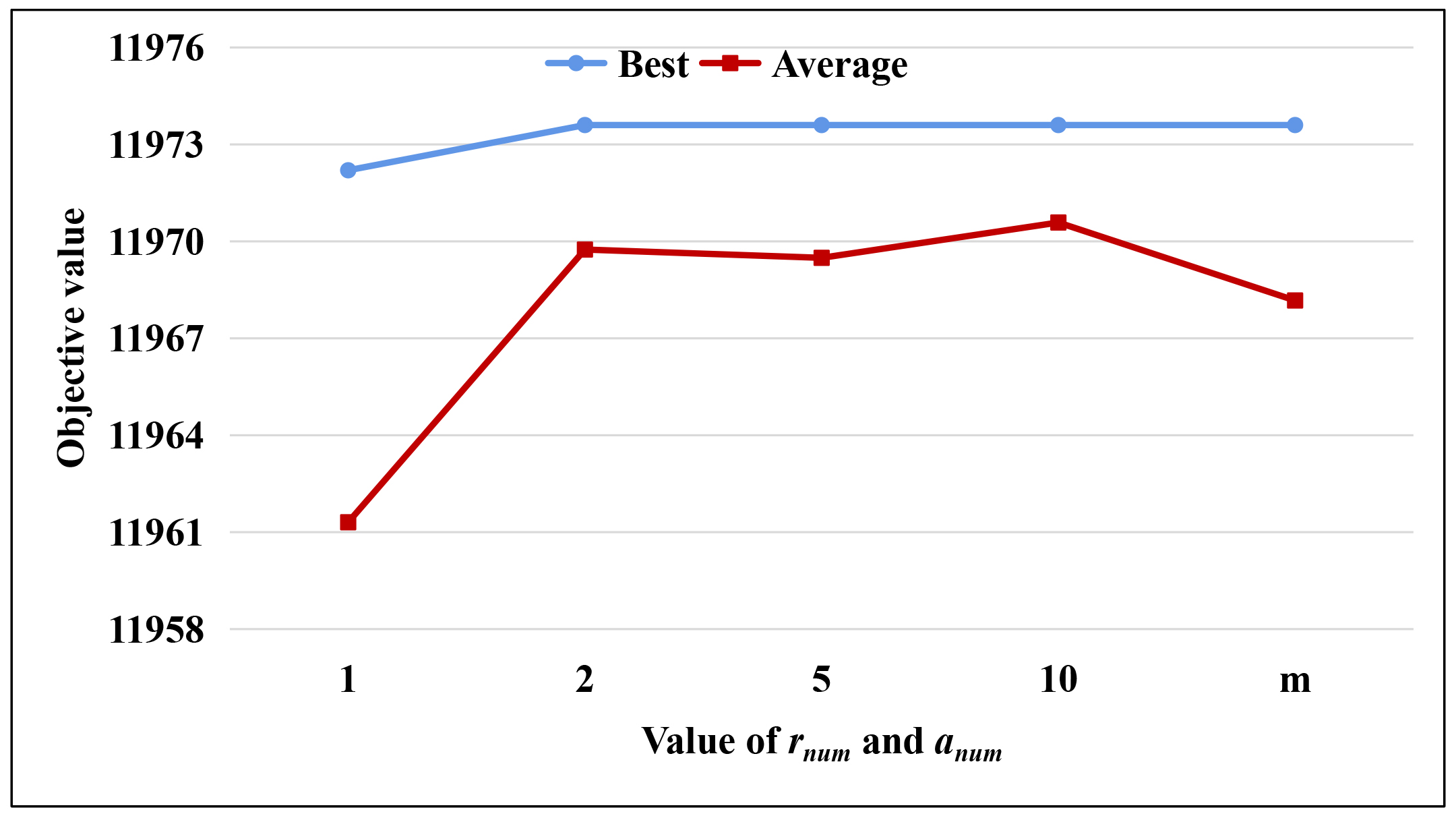}  
\label{fig_Para-I}}
\subfigure[Comparison results on \textit{Set A}]{
\includegraphics[width=0.96\columnwidth]{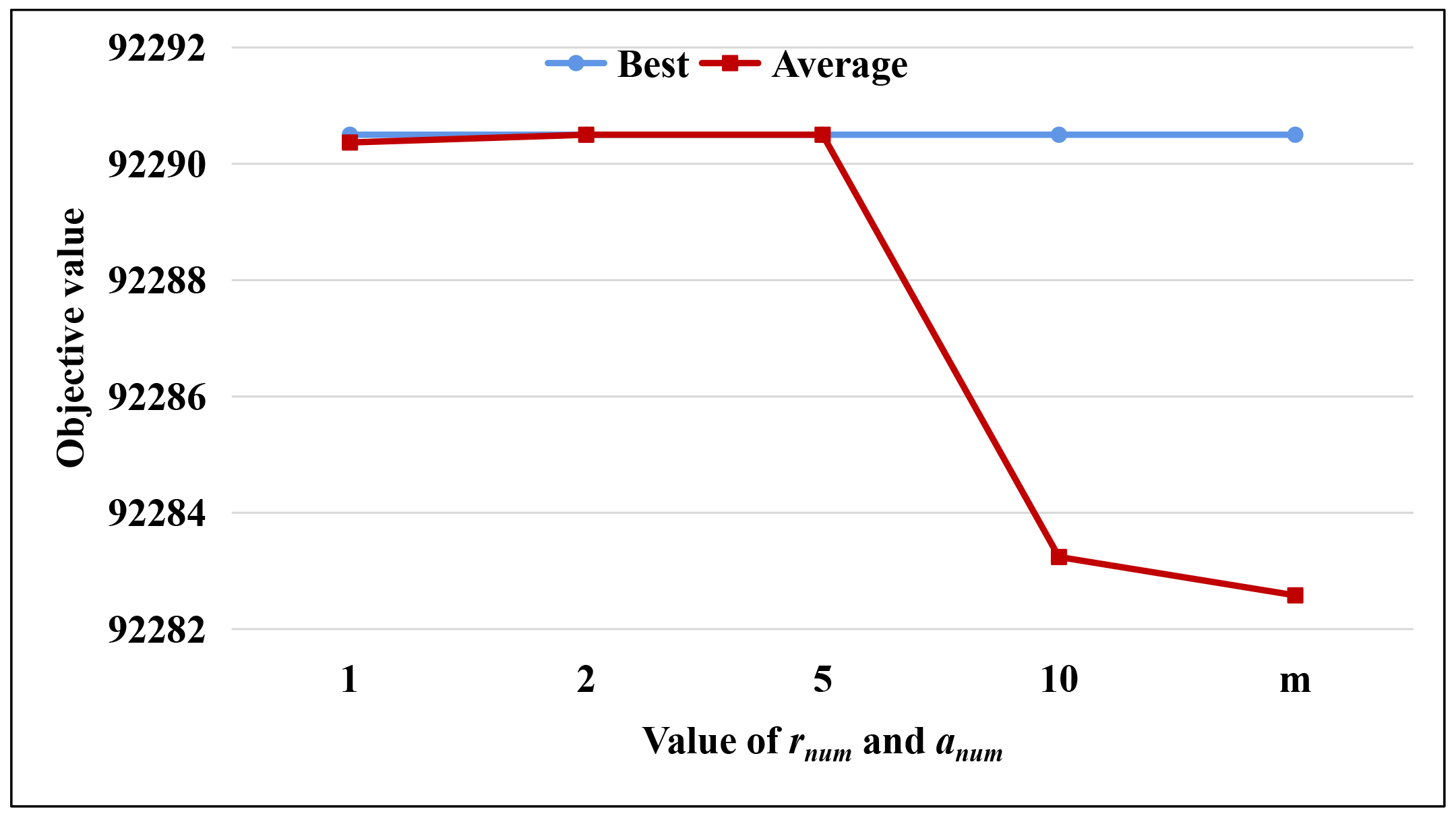}  
\label{fig_Para-A}}
\subfigure[Comparison results on \textit{Set II}]{
\includegraphics[width=0.96\columnwidth]{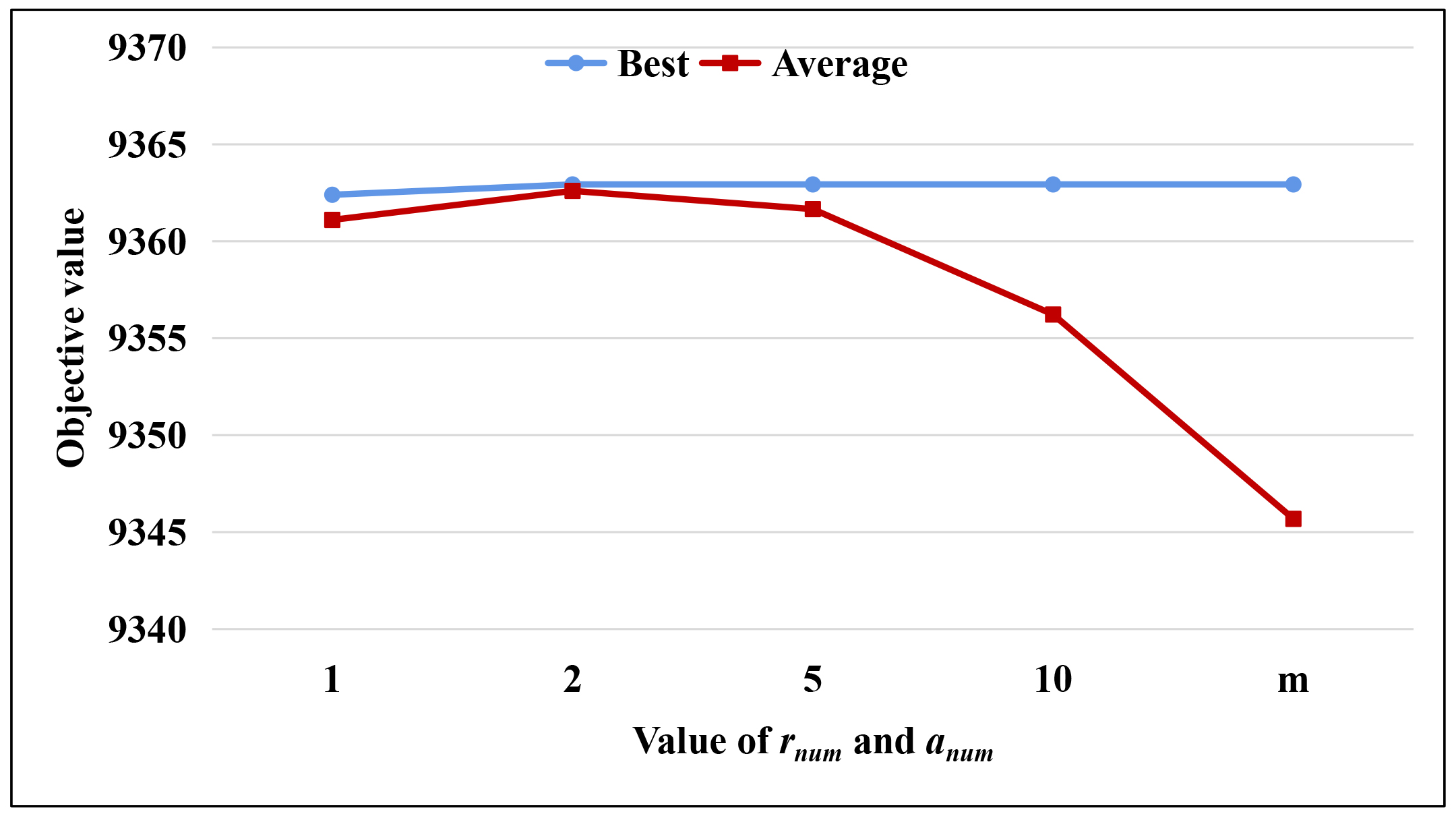}  
\label{fig_Para-II}}
\subfigure[Comparison results on \textit{Set B}]{
\includegraphics[width=0.96\columnwidth]{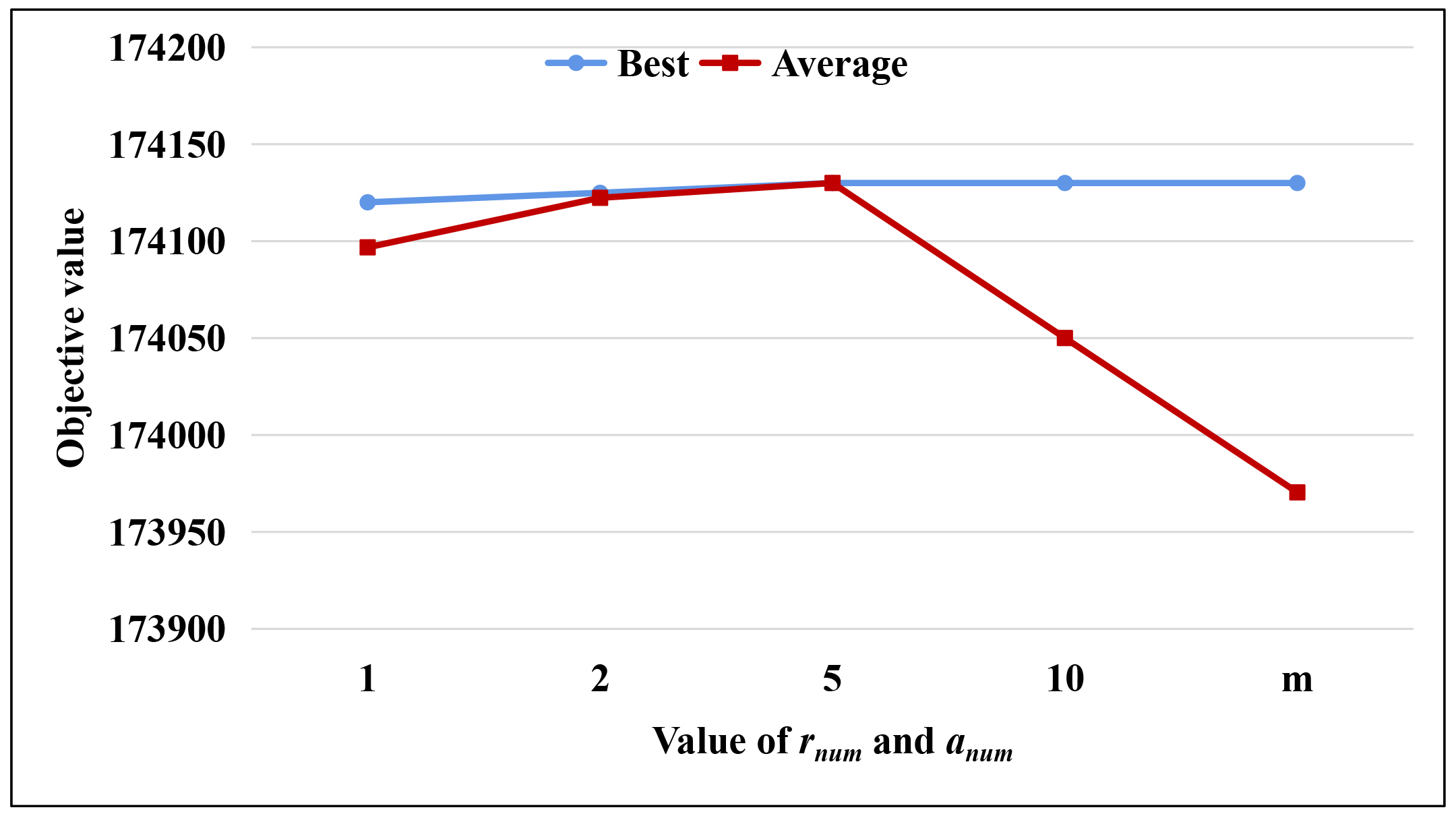}
\label{fig_Para-B}}
\subfigure[Comparison results on \textit{Set III}]{
\includegraphics[width=0.96\columnwidth]{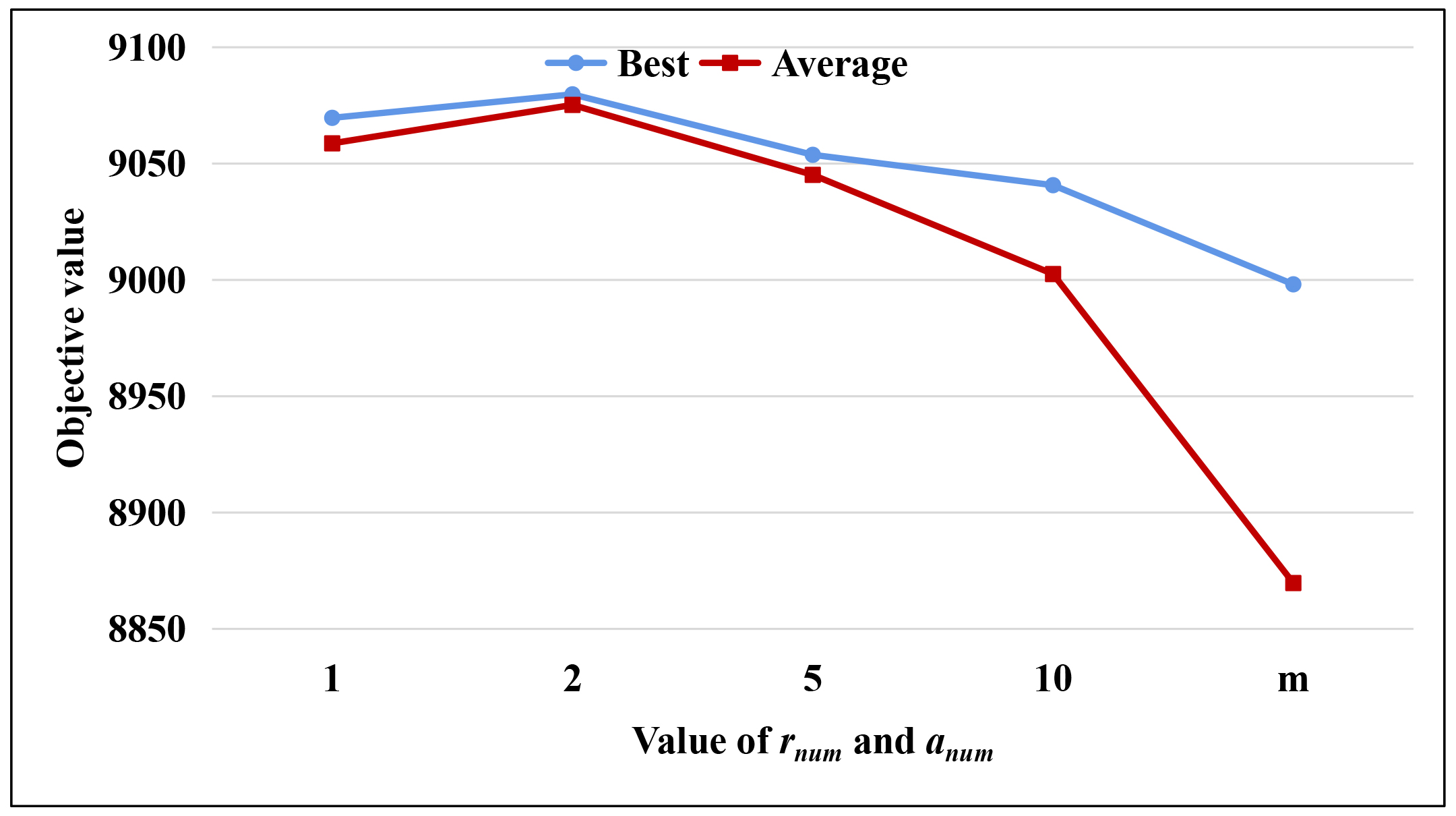}
\label{fig_Para-III}}
\subfigure[Comparison results on \textit{Set C}]{
\includegraphics[width=0.96\columnwidth]{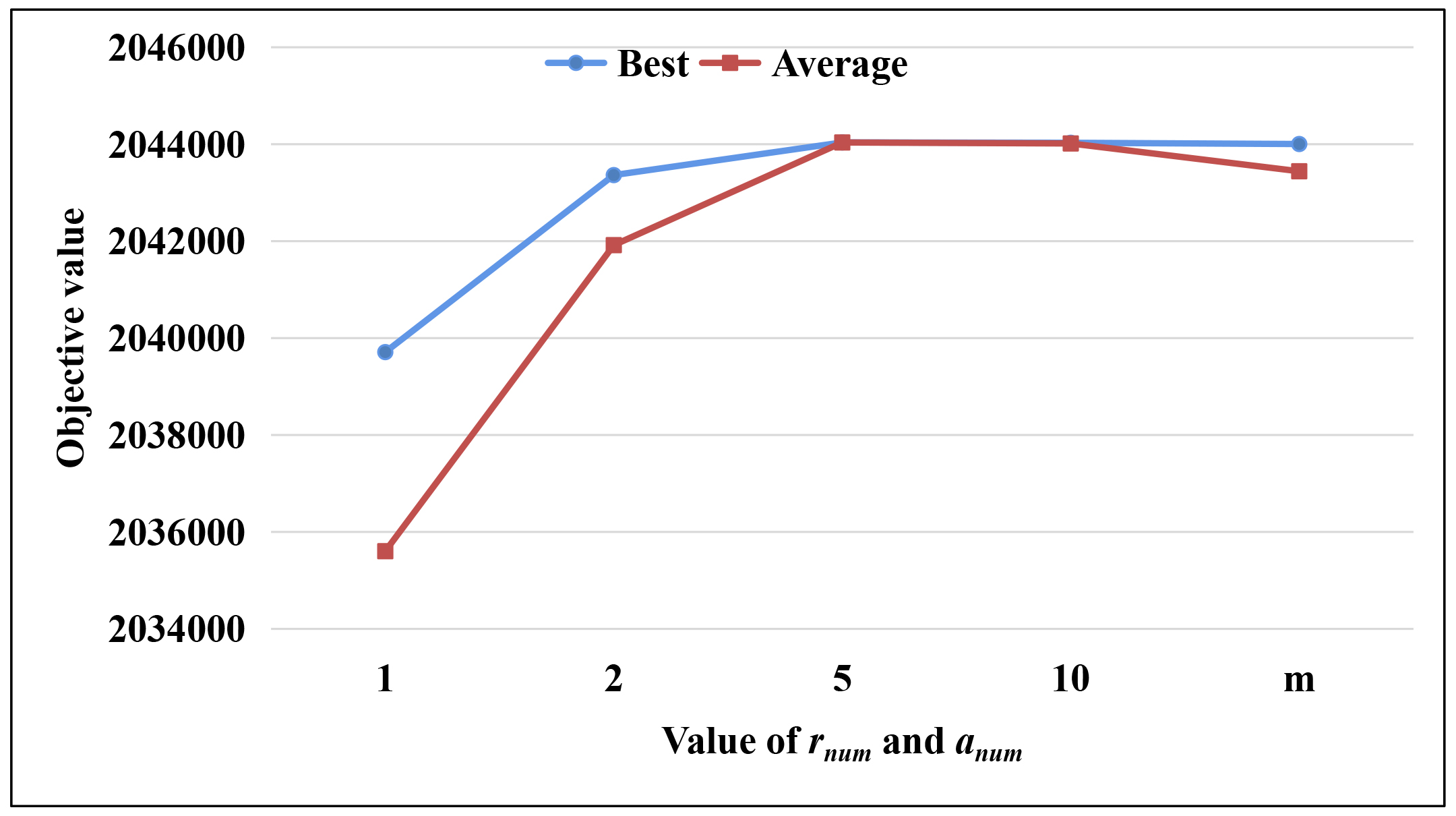}
\label{fig_Para-C}}
\caption{Analyses on the influence of parameters $r_{num}$ and $a_{num}$ on the performance of E2LS. The results are expressed by the average values of the best and average solutions in 10 runs obtained by E2LS with different settings of parameters on all the instances in each instance set.}
\label{fig_Para}
\end{figure*}

\begin{table*}[t]
\caption{Comparisons of E2LS, E2LS($t=1$), E2LS($t=10$), and the baseline algorithms on three sets of SUKP instances. The results are expressed by the average values of the best and average solution, standard deviations, and average run times obtained by each algorithm on all the instances in \textit{Sets II} and \textit{III}.}
\label{table-SUKP-para}
\centering
\scalebox{0.78}{\begin{tabular}{lrrrrrrrrrrrrrrr} \bottomrule
Algorithm\textbackslash{}Benchmark &  & \multicolumn{4}{r}{\textit{Set I}}    &  & \multicolumn{4}{r}{\textit{Set II}} &  & \multicolumn{4}{r}{\textit{Set III}} \\ \cline{3-6} \cline{8-11} \cline{13-16} 
                                   &  & Best     & Average  & S.D.  & Time    &  & Best    & Average & S.D.  & Time    &  & Best    & Average & S.D.   & Time    \\ \hline
I2PLS                              &  & 11964.67 & 11938.10 & 25.90 & 90.740  &  & 9274.47 & 9210.91 & 45.96 & 333.210 &  & 8583.50 & 8550.47 & 27.91  & 297.920 \\
KBTS                               &  & 11973.60 & 11969.45 & 5.81  & 59.714  &  & 9348.67 & 9307.43 & 21.32 & 243.472 &  & 9024.28 & 9005.18 & 23.14  & 147.623 \\
MSBTS                              &  & 11970.03 & 11955.78 & 18.35 & 240.734 &  & 9361.87 & 9352.69 & 5.78  & 187.351 &  & 8748.78 & 8619.14 & 130.33 & 104.053 \\
ATS-DLA                            &  & 11932.47 & 11864.17 & 49.69 & 11.693  &  & 9358.80 & 9357.36 & 0.48  & 39.951  &  & 9048.89 & 9033.42 & 9.87   & 236.289 \\
E2LS($t = 1$)                        &  & 11973.60 & 11969.08 & 4.80  & 48.581  &  & 9362.93 & 9362.55 & 0.15  & 40.110  &  & 9079.33 & 9074.93 & 5.54   & 235.147 \\
E2LS($t = 10$)                       &  & 11973.60 & 11969.49 & 6.21  & 43.251  &  & 9362.93 & 9362.59 & 0.19  & 37.664  &  & 9079.61 & 9075.19 & 4.70   & 237.549 \\
E2LS                               &  & 11973.60 & 11969.75 & 5.60  & 42.173  &  & 9362.93 & 9362.60 & 0.19  & 40.235  &  & 9079.78 & 9075.23 & 4.46   & 227.603  \\ \toprule
\end{tabular}}
\end{table*}
\begin{table*}[t]
\caption{Comparisons of E2LS, E2LS($t=1$), E2LS($t=10$), and the baseline algorithms on three sets of BMCP instances. The results are expressed by the average values of the best and average solution obtained by each algorithm, coupled with the average values of the standard deviations and average run times obtained by E2LS, on all the instances in \textit{Sets B} and \textit{C}.}
\label{table-BMCP-para}
\centering
\scalebox{0.75}{\begin{tabular}{lrrrrrrrrrrrrrrr} \bottomrule
Algorithm\textbackslash{}Benchmark &  & \multicolumn{4}{r}{\textit{Set A}} &  & \multicolumn{4}{r}{\textit{Set B}}    &  & \multicolumn{4}{r}{\textit{Set C}}        \\ \cline{3-6} \cline{8-11} \cline{13-16} 
                                   &  & Best     & Average  & S.D. & Time  &  & Best      & Average   & S.D. & Time   &  & Best       & Average    & S.D.  & Time    \\ \hline
PLTS                               &  & 91533.67 & 91276.79 & --   & --    &  & 172471.63 & 171196.07 & --   & --     &  & 1967744.97 & 1943665.33 & --    & --      \\
VDLS                               &  & 92290.50 & 91980.88 & --   & --    &  & 174048.87 & 173212.76 & --   & --     &  & 2042027.00 & 2033356.73 & --    & --      \\
E2LS($t = 1$)                        &  & 92290.50 & 92290.50 & 0.00 & 2.508 &  & 174130.00 & 174130.00 & 0.00 & 20.458 &  & 2044036.53 & 2044033.62 & 8.75  & 100.576 \\
E2LS($t = 10$)                       &  & 92290.50 & 92290.50 & 0.00 & 2.792 &  & 174130.00 & 174130.00 & 0.00 & 36.327 &  & 2044036.53 & 2044030.43 & 18.31 & 81.638  \\
E2LS                               &  & 92290.50 & 92290.50 & 0.00 & 2.676 &  & 174130.00 & 174130.00 & 0.00 & 30.668 &  & 2044036.53 & 2044035.93 & 1.81  & 70.037 \\ \toprule
\end{tabular}}
\end{table*}

\subsection{Analyses on the parameters}
\label{Sec_para}

We then analyze the influence of the parameters including $r_{num}$, $a_{num}$, and $t$ on the performance of E2LS by comparing E2LS with its variants on each instance set. 
We first compare the E2LS algorithm with different values of $r_{num}$ and $a_{num}$. We tested five pairs of parameters. They are $r_{num},a_{num} = 1,2,5,10,m$ respectively. The results are shown in Figure \ref{fig_Para}, that compares the average values of the best and average solutions of each algorithm on all the instances in each instance set.

The results in Figure \ref{fig_Para} show that with the increase of the parameter values from 1 to $m$, the performance of E2LS first increases and then decreases. This result indicates that balancing the search efficiency and the search region is reasonable and necessary. When we set small values to the parameters (e.g., 1 for the SUKP, 1 or 2 for the BMCP), E2LS can explore the search space very quickly, but it might be easy to get stuck in some local optima. When we set large values to the parameters (e.g., 10 or $m$), E2LS can explore the search space widely and deeply, but the low efficiency might also make it hard to escape from local optima.

We can also observe that the properties of the SUKP and BMCP are different. For the SUKP, E2LS with small values of parameters are not good at solving the small instances in \textit{Set I}, and E2LS with large values of parameters are not good at solving the large instances in \textit{Sets II} and \textit{III}. This result is consistent with the results of the SUKP baseline algorithms, and can explain their performance. That is, algorithms KBTS and MSBTS with large and unrefined search regions are good at solving the small instances in \textit{Set I}, but do not work well on \textit{Set III}. In contrast, ATS-DLA with a small and overrefined search region works well on \textit{Sets II} and \textit{III}, but not on \textit{Set I}.

For the BMCP, the instances in \textit{Sets A} and \textit{B} prefer small parameter values to large ones. While the instances in \textit{Set C} prefer large parameter values to small ones. This might be because for the relatively small instances in \textit{Sets A} and \textit{B} (with 585 to 1,600 items or elements), low search efficiency is more likely to get stuck in local optima than a small search region. While the situation for the large instances in \textit{Set C} (with 4,000 to 5,200 items or elements) is on the contrary. 


Moreover, the best settings of parameters when solving the SUKP and BMCP are different. The best settings of $r_{num}, a_{num}$ are 2 and 5 for the SUKP and BMCP, respectively. This is also because of the different properties of these two problems. For example, when solving the SUKP, items with zero additional weights will not be added into the feasible candidate set $FC$ (which will be refined by the parameter $a_{num}$), and will be added into the current solution directly by the ADD$^*$ operator if such a move is not prohibited by the tabu list (lines 5-8 in Algorithm \ref{alg_add}). While when solving the BMCP, there is no item with zero additional weights (values), since the value of each item is positive (in the BMCP, an item with zero values should always be contained in the solution). Therefore, if we set the parameter $a_{num}$ to be the same when solving the SUKP and BMCP, the ADD$^*$ operator for the SUKP can add more items than that for the BMCP. Thus it is reasonable to set larger parameter values for the BMCP than for the SUKP.

We further analyze the influence of the parameter $t$ on the performance of E2LS. We denote E2LS($t = 1$) and E2LS($t = 10$) as two variants of E2LS with the parameter $t$ equals to 1 and 10, respectively. Note that E2LS($t = 1$) actually generates the initial solution randomly. We compare these two variants with E2LS and the baseline algorithms. Tables \ref{table-SUKP-para} and \ref{table-BMCP-para} show the results of the algorithms for the SUKP and BMCP, respectively. The results are the average values of the best and average solutions, the standard deviations, and the average run times of each algorithm on all the instances in each instance set.

As the results show, both E2LS($t = 1$) and E2LS($t = 10$) can obtain competitive results with E2LS, and their performance is significantly better than the baseline algorithms for both the SUKP and BMCP. The result indicates that the local search method in E2LS has excellent stability and robustness, as it can yield excellent results with various initial solutions, even if they are generated randomly. Moreover, E2LS slightly outperforms E2LS($t = 1$) and E2LS($t = 10$), indicating that higher-quality initial solutions lead to better performance, and the random greedy initialization method in E2LS is effective.


\section{Conclusion}
\label{Sec_Conclusion}
This paper proposes an efficient and effective local search algorithm called E2LS for the SUKP and BMCP problems. To our knowledge, this is the first time that an algorithm has been proposed for these two closely related problems. The E2LS algorithm can explore the solution space efficiently by refining the search region, i.e., abandoning the low-quality moves. The proposed ADD$^*$ operator in E2LS can traverse the refined search region and provide high-quality moves quickly. As a result, E2LS trade-offs the search region and efficiency well, which leads to an excellent performance. Such a trade-off mechanism and the approach of traversing the refined search region could be applied to various combinatorial optimization problems.

Extensive experiments on 78 public SUKP instances and 90 public BMCP instances with various scales demonstrate the superiority of the proposed algorithm. In particular, E2LS provides four new best-known solutions for the SUKP, and 27 new best-known solutions for the BMCP.


\section*{Declarations of interest}
None.


\bibliographystyle{cas-model2-names}
\bibliography{main}


\end{document}